\documentclass[english]{lni}

\IfFileExists{latin1.sty}{\usepackage{latin1}}{\usepackage{isolatin1}}

\usepackage{graphicx}
\usepackage{multirow}

\title{Knowledge and Metadata Integration\\for Warehousing Complex Data}

\author{
	Jean-Christian Ralaivao $^{1,2}$ and J\'{e}r\^{o}me Darmont $^2$\\ \\
	\begin{tabular}{cc}
	$^1$ \'{E}cole Nationale d'Informatique & $^2$ Universit\'{e} de Lyon (ERIC Lyon 2) \\
	BP 1487 & 5 avenue Pierre Mendès-France \\
	Fianarantsoa (301) & 69676 Bron Cedex \\
	Madagascar & France \\
	ralaivao@mail.univ-fianar.mg & first\_name.last\_name@eric.univ-lyon2.fr \\
	\end{tabular}
}

\begin{document}
\maketitle

\begin{abstract}
With the ever-growing availability of so-called complex data, especially on the Web, decision-support systems such as data warehouses must store and process data that are not only numerical or symbolic. Warehousing and analyzing such data requires the joint exploitation of metadata and domain-related knowledge, which must thereby be integrated. In this paper, we survey the types of knowledge and metadata that are needed for managing complex data, discuss the issue of knowledge and metadata integration, and propose a CWM-compliant integration solution that we incorporate into an XML complex data warehousing framework we previously designed.
\end{abstract}

\section{Introduction}
\label{sec:Introduction}

Decision-support technologies, and more particularly data warehousing \cite{inmon02,kimball02,jarke03}, are nowadays technologically mature. Data warehouses are aimed at monitoring and analyzing activities that are materialized by numerical measures (facts), while symbolic data describe these facts and constitute analysis axes (dimensions). However, in real life, many decision-support fields (customer relationship management, marketing, competition monitoring, medicine...) need to exploit data that are not only numerical or symbolic. For example, computer-aided diagnosis systems might require the analysis of various and heterogeneous data, such as patient records, medical images, biological analysis results, and previous diagnoses stored as texts \cite{saad04}. We term such data \emph{complex data} \cite{iceis05}. Their availability is now very common, especially since the broad development of the Web, and more recently the Web 2.0 (blogs, wikis, multimedia data sharing sites...).

Complex data might be structured or not, and are often located in several, heterogeneous data sources. Specific approaches are needed to collect, integrate, manage and analyze them. A data warehousing solution is interesting in this context, though adaptations are obviously necessary to take into account data complexity (measures might not be numerical, for instance). Data volumetry and dating are also other arguments in favor of the warehousing approach.

In this context, metadata and domain-related knowledge are essential in the processing of complex data and play an important role when integrating, managing, and analyzing them. In this paper, we address the issue of jointly managing knowledge and metadata, in order to warehouse complex data and handle them, at three different levels:
     at the supplier level (data providers), to identify all input data sources and the role of source type drivers;
     at the user level (consumers), to identify all data sources for analysis and their source type drivers;
     at the manager level (administrators), to achieve good performance.
%\begin{itemize}
    %\item at the supplier level (data providers), to identify all input data sources and the role of source type drivers;
    %\item at the user level (consumers), to identify all data sources for analysis and their source type drivers;
    %\item at the manager level (administrators), to achieve good performance.
%\end{itemize}

Since data warehouses traditionally handle knowledge under the form of metadata, we discuss the alternatives for integrating domain-related knowledge and metadata. Our position is that knowledge should be integrated as metadata in a complex data warehouse. On this basis, we also present an XML-based architecture framework for complex data warehouses that expands the one we proposed in \cite{iceis05}.

The remainder of this paper is organized as follows. In
Section~\ref{sec:KnowledgeAndMetadataNeeds}, we survey the
various kinds of knowledge and metadata that are required for
managing complex data. In
Section~\ref{sec:KnowledgeAndMetadataIntegrationForComplexDataWarehousing},
we discuss the issue of knowledge and metadata integration,
justify our choice, and present our revised architecture framework
for complex data warehouses. In Section~\ref{sec:StateOfTheArt},
we summarize the state of the art regarding knowledge and metadata integration. We
finally conclude this paper and provide future research directions
in Section~\ref{sec:Conclusion}.

\section{Knowledge and Metadata Needs}
\label{sec:KnowledgeAndMetadataNeeds}

\subsection{Knowledge Types}
\label{sec:KnowledgeTypes}

Two types of knowledge must be taken in consideration: tacit and
explicit knowledge \cite{kw02}. Tacit knowledge includes beliefs,
perspectives and mental models. Explicit knowledge is knowledge
that can be expressed formally using a language, symbols, rules,
objects or equations, and thus can be communicated to others. In
data warehousing environments, we are particularly interested in
explicit knowledge.

Then, different kinds of questions must be considered regarding the types of
knowledge that are needed to manage complex data warehouses. These
questions determine the description context (what), the
organizational context (who, where and when), the processing
context (how) and the motivation and business rules (why).

Responses to the ``what"-type question describe 
business concepts. These elements guide the link between metadata and
knowledge; while knowledge representation uses metadata contents and
structure.
The ``how" and ``why" questions relate to each process' motivation and the way it operates, in comparison to an existing organization.
Eventually, answering to the ``who", ``where" and ``when" questions helps in connecting
the first two categories of questions to a particular
organization.

%- Connaissance universelle (de fond)
%    Exemples : Nb de jours par semaine, par mois, par an, Calendrier (jours fériés, jours ouvrés, …)

Furthermore, one type of knowledge that is often forgotten is universal or
background knowledge. For example, the number of days in a month,
the work scheduler with wrought days, public holidays, constitute
some background knowledge that is essential for decision or
analytical queries.

%- Connaissance statistique sur le contenu + échantillon
%    Exemples : statistiques descriptives sur les données,
%    hypothèses sur les données (lois, paramètres, ...), Summaries

We must also consider statistical knowledge, which may include
descriptive statistics about the data warehouse contents, or
hypotheses about attributes' characteristics, such as probabilistic
laws or sampling methods. Statistical knowledge may be provided
by data analysis or data mining, and results should be reinjected into the system.

%- Connaissances techniques
%    - Métadonnées : Sources des données, types de données
%    - Types de systèmes de gestion de données (SBGD disponibles, plateforme, technologie, …)
%    - Techniques d'indexation, possibilités offertes par chaque
%    SGBD, possibilités pour chaque type de données

Technical knowledge is also very important at different phases of the data warehouse lifecycle. At a high level of
abstraction, it is closely related to metadata. Technical knowledge includes knowledge about data sources and targets,
standard and specific data types, database management systems
(DBMSs), software and hardware platforms, technologies, etc.  Indexing techniques available in
each DBMS belong to this type of knowledge too. 

%- Connaissance sur l'organisation
%    - sites organisationnels et déploiement du système
%    - utilisateurs, besoins des utilisateurs, contraintes par
%    rapport aux besoins (période, temps de réponse, volume de
%    données traité, format des résultats attendus, ...)

Closely related is knowledge about organizational and geographical deployment, which 
includes information about users, their needs, their attributions and their
constraints in regard to their needs (e.g., in terms of response time, volume of processed data, result format, etc.).

%- Connaissance sur l'administration de l'entrepôt
%    - Statistique d'accès
%    - Connaissance sur les applications OLTP (fréquences, temps de
%réponse, utilisateurs concernés, ...)
%    - ETL :planification, calendrier, période de pointe, …
%    - Rafraichissement de l'entrepôt : Critère de regroupement -
%Période de rotation des données récapitulatives - Période de purge
%/ calcul des données dormantes

The last kind of knowledge we must consider relates to data
warehouse administration. It provides information about how the
data warehouse is used (access statistics) and how the interface
between the data warehouse and the operational systems
articulates, i.e., what the transactional applications and their
characteristics (frequencies, response times, users...) are; and what
the major Extracting, Transforming and Loading (ETL) problems  
(planification to satisfy user requirements with respect
to work schedule, identification of peak periods...) are.
The refreshment policies of the data warehouse contents are also
important here, since they dictate the rotation period of summary
data, the purge period and dormant data determination.

\subsection{Metadata Types}
\label{sec:MetadataTypes}

% Partie 4 de ton document, a rediger.

% [Metadata classifications in Metadata for Object-Relational Data
% Warehouse p.3-4]:

We identified five transversal and complementary classifications for metadata in the
literature.
In the first classification \cite{mordw00}, metadata are
classified based on the data warehouse architecture layers, as
follows:
\begin{itemize}
    \item metadata associated with data loading and transformation, which describe the source data and any changes operated on
    data;
    \item metadata associated with data management, which define the data stored in the data
    warehouse;
    \item metadata used by the query manager to generate an appropriate
    query.
\end{itemize}

The second classification \cite{mordw00} divides metadata into:
\begin{itemize}
    \item technical metadata that support the technical staff and contain the terms and definition of metadata as they appear in operational databases;
    \item business metadata that support business end-users who do not have any technical background;
    \item information navigator metadata, which are tools that help users navigate through both the business metadata and the warehoused data.
\end{itemize}

In the third classification \cite{mordw00}, metadata may be:
\begin{itemize}
    \item static metadata that are used to document or browse the system;
    \item dynamic metadata that can be generated and maintained at run time. A new kind of metadata is made of metadata that handle the mapping between systems.
\end{itemize}

In the fourth classification \cite{ommt05}, metadata may be:
\begin{itemize}
    \item system catalog metadata or data descriptors;
    \item relationship metadata that store information about the relationships between data entities (primary key/foreign key relationships, generalization/specialization relationship,
aggregation relationship, inheritance relationships and any other
special semantic relationship implying update or delete
dependency);
    \item content metadata formed by descriptions of the contents of stored data at an arbitrary granularity. Content metadata may be as
simple as one keyword, or as complex as a business rules, formulae
or links to whole documents;
    \item data lineage metadata, which are lifecycle data about stored data (information about the creation of data, subsequent updates, transformation, versioning,
summarization, migration, and replication, transformation rules,
and descriptions of migration and replication);
    \item technical metadata that store technical information about stored data: format, compression or encoding algorithm used, encryption and decryption
algorithms, encryption and decryption keys, software used to
create or update the data, Application Programming Interfaces (APIs) available to access the data, etc.;
    \item data usage metadata or business data that are descriptions of how and for what purposes the data are to be used by users and applications;
    \item system metadata that are descriptions about the overall system environment, including hardware, operating system and application
software;
    \item process metadata that describe the processes in which the applications operate, and any relevant output of each step of these processes.
\end{itemize}

Eventually, the fifth classification we identified \cite{mmde06}
is based upon functionality categories: infrastructure, data
model, process, quality, interface and administration.
\begin{itemize}
	\item Infrastructure metadata contain information on system components.
	\item Data model metadata (also called data dictionary) include
definitions of data entities and the relationships among them.
	\item Process metadata capture information on data generation and
transfer from sources to targets. 
	\item Quality metadata contains
information on the actual data stored and helps in assessing data
quality (e.g., factual measurements). 
	\item Interface metadata (also
called reporting metadata) support data delivery to end-users.
	\item Finally, administration metadata include data that are necessary for
administering the data warehouse and its associate applications
(security, authentification, usage tracking...).
\end{itemize}

%[A Software Architecture for XML-based Metadata Interchange in
%Data Warehouse Systems p.7]:
%
%A central metadata repository implements a common metamodel based
%on CWM and serves as hub for metadata interchange. Local metadata
%includes: - Business Intelligence Metadata (OLAP, Data Mining,
%Reporting, ad hoc business applications) - Derived Data Storage
%Metadata (Data Mart) - Data Movement Metadata (Aggregate, Select,
%Load - Extract, Clean, Transform, Load) - Data Warehouse MetaData
%- Data Sourcing Metadata (Operational Data Sources, External Data
%Sources)

To conclude this section, we cite an important standardization initiative: the Common Warehouse Metamodel
 (CWM \cite{cwm01}). CWM has been established by the Object Management Group
(OMG) within its framework of Meta-Object Facilities (MOF). CWM
purposes a metamodel that can be instantiated to obtain an
operational data warehouse. Each of the metadata types we enumerated in the above
classifications should be mapped into one or several CWM
components.

\section{Knowledge and Metadata Integration for Complex Data Warehousing}
\label{sec:KnowledgeAndMetadataIntegrationForComplexDataWarehousing}

\subsection{Integrating Knowledge and Metadata}
\label{sec:IntegratingKnowledgeAndMetadata}

% Partie 5 de ton document, a rediger.

%\subsubsection{Why Integration?}

Current data warehouse architectures are based on metadata.
However, they are sometimes themselves a materialization of domain-related
knowledge that facilitates the management of data warehouses and
helps in achieving good performance. It is difficult for classical
architectures to manage complex data without domain-related knowledge nor
background knowledge. For example, a data warehouse administrator
needs some background, domain-related knowledge in addition to metadata to
select clustering or indexing techniques.

%\subsubsection{How to Operate the Integration?}
%\label{choice}

There are three possibilities to jointly manage knowledge and metadata: 
%\begin{enumerate}
	 coding and representing knowledge as metadata;
	 modelling metadata to match knowledge representation;
	 managing metadata and knowledge separately.
%\end{enumerate}
The advantages and drawbacks of each possibility are discussed below.

Coding and representing knowledge as metadata present an important
advantage: we can keep on using and maintain current architectures and
techniques. However, it is necessary to find a solution for
knowledge representation, a kind of mapping between classical
knowledge representation and metadata implementation.

Modelling metadata to match knowledge representation hedges on the
domain of knowledge warehouses \cite{kw02}, which supposes
important adaptations and new considerations about current architectures. Some
metadata cannot be converted into knowledge and there is a risk to
loose some information. Moreover, finding a knowledge
representation that can accept actual metadata is not obvious.

As for the third possibility, i.e., managing metadata and
knowledge separately, a great change of architecture would be
essential, because a structure that allows to coordinate and to
compile metadata and knowledge contents must be devised. Instead
of reducing complexity, this solution would increase it with the
consideration of a new element: managing the connection between
knowledge and metadata.

In conclusion, in order to build upon the assets of current data warehouse
architectures, in particular in terms of performance, we select the first
solution and explore it in this paper.

\subsection{Revised Architecture Framework for Complex Data Warehousing}
\label{sec:revisedArchitectureFrameworkForComplexDataWarehousing}

% Partie 2 de ton document, a rediger.

\subsubsection{Global Architecture}
\label{globalarch}

\begin{figure}[hbt]
{\centering
\resizebox*{1.0\textwidth}{!}{\includegraphics{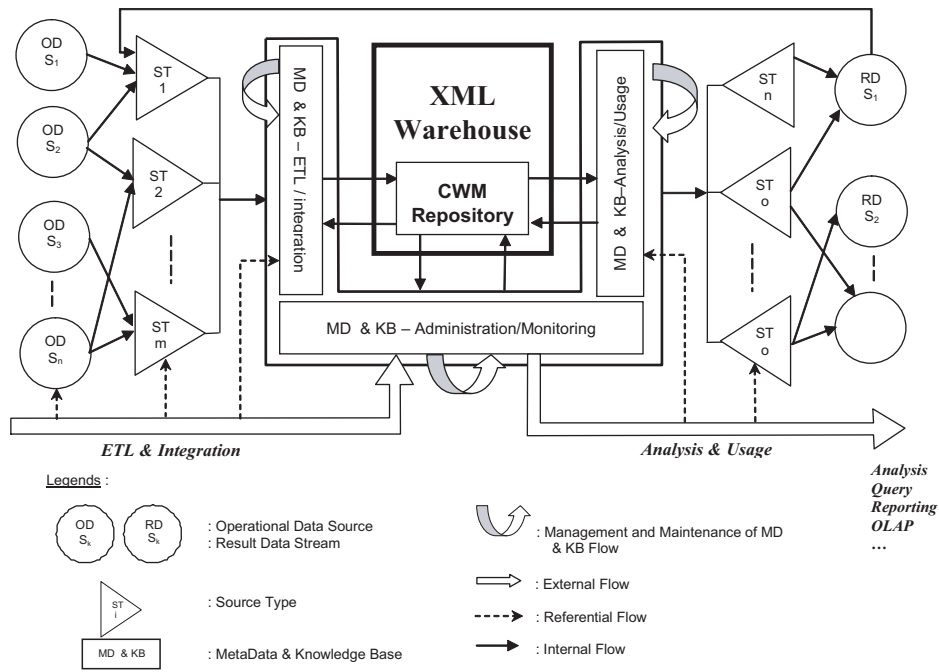}}
\par}
\caption{Complex Data Warehouse Architecture Framework}
\label{fig:Figure1}
\end{figure}

In \cite{iceis05}, we have already proposed an architecture framework for complex data
warehouses. The main components of this architecture (Figure~\ref{fig:Figure1}) are:
the data warehouse kernel, which may be either materialized as an XML warehouse, or virtual (where cubes are computed at run time);
operational databases;
source type drivers that notably include mapping specifications between the sources and XML;
and finally a metadata and knowledge base layer that includes three submodules related to three management processes.
%\begin{itemize}
%    \item the data warehouse kernel, which may be either materialized as an XML warehouse, or virtual (where cubes are computed at run time);
%    \item operational databases;
%    \item source type drivers that notably include mapping specifications between the sources and XML;
%    \item a metadata and knowledge base layer that includes three submodules related to three management processes.
%\end{itemize}

These three processes for managing a data warehouse are:
\begin{enumerate}
    \item the ETL and integration process that feeds the warehouse with source data from the operational databases ($OD$) by using drivers that are specific to each source type ($ST$);
    \item the administration and monitoring process ($MD~\&~KB$) that manages metadata and knowledge (the administrator interacts with the data warehouse through this process);
    \item the analysis and usage process that runs user queries, produces reports, builds data cubes, supports On-Line Analytical Processing (OLAP), etc. (result data $RD$).
\end{enumerate}

Each of these processes exploits and updates the metadata and the
knowledge base through four types of flows:
\begin{enumerate}
    \item the external flow, which includes the ETL and integration flow and the exploitation (analysis and usage) flow (the warehouse may thus be considered as a black box);
    \item the internal flow, between the warehouse kernel and the metadata and knowledge base layer, and between the metadata and knowledge base layer and the source type drivers;
    \item the metadata and knowledge management and maintenance flow, which acquires new knowledge and enriches existing knowledge;
    \item the reference flow, which illustrates the fact that the external flow always refers to the metadata and knowledge base layer for integration, ETL, and analysis and usage in general.
\end{enumerate}

The symmetric aspect between ``sources" and ``usages" around the
data warehouse core allows us to eventually re-inject results as
data sources. % à développer
For instance, a data mining analysis may discover dependencies between variables and highlight causal relationships among them. We do use such techniques to determine the relevance of complex data with respect to given analysis goals. 
Then, knowledge obtained by mining can be integrated into the metadata repository and later re-used in the definition of complex data cubes. 

\subsubsection{Core Interface}

\begin{figure}[hbt]
{\centering
\resizebox*{0.5\textwidth}{!}{\includegraphics{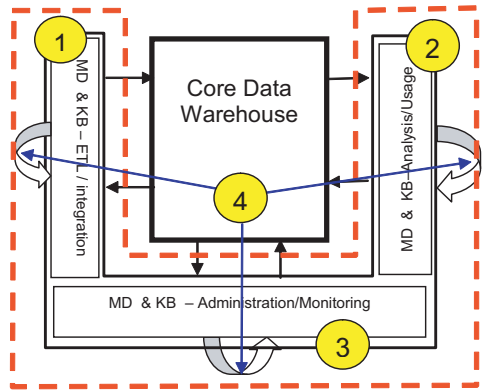}}
\par}
\caption{Interface around the core}
\label{fig:Figure2}
\end{figure}

In this section, we expand the architecture framework presented in Section~\ref{globalarch}
by integrating knowledge and metadata.
Around the data warehouse core, with respect to the external
components (operational data sources, result data stream and
administration and monitoring), we define three metadata and
knowledge base ($MD~\&~KB$) repositories corresponding to the
three sides of the core (Figure~\ref{fig:Figure2}). They constitute an interface functionality.

The first $MD~\&~KB$ repository (labeled (1) in Figure~\ref{fig:Figure2}) lies at the data integration and ETL process
level, and includes:
\begin{itemize}
    \item an ontology for modelling domain-related knowledge;
    \item information about data sources and source types;
    \item mappings for the extraction and transformation processes (the E and T in ETL);
    \item information about the loading (the L in ETL: frequency, mode...) and cleansing (purge) processes;
    \item a referential or metadata repository about data, materialized views, index, clusters, aliases, etc.
\end{itemize}

The $MD~\&~KB$ repository that is labeled (3) in Figure~\ref{fig:Figure2} lies at the administration and monitoring level, and
references :
\begin{itemize}
    \item an ontology for modelling domain-related knowledge;
    \item deployment, hardware and software constraints;
    \item an interface between the integration and ETL level and the usage level;
    \item information on users and data providers;
    \item data warehouse usage information (statistics, response time, availability, feedback, dormant data...);
    \item a referential or metadata repository about data, materialized views, index, clusters, aliases, etc.
\end{itemize}

Eventually, the $MD~\&~KB$ that is labeled (2) in Figure~\ref{fig:Figure2} lies at the usage level and completes our interface with:
\begin{itemize}
    \item an ontology for modelling domain-related knowledge;
    \item information about aggregate operators (hierarchical lattice construction \cite{pei03} if necessary) and
    data lineage that would allow users to go up to the sources if
    necessary;
    \item query optimizer data (query reformulation and rewriting);
    \item a referential or metadata repository about data, materialized views, index, clusters, aliases, etc.
\end{itemize}

Note that some of the elements we have just enumerated (e.g., ontology and referential repository) are
present in more than one interface. Hence, they must be factorized at a higher level (labeled (4) in Figure~\ref{fig:Figure2}).
Moreover, this level must include metaknowledge, i.e.,
knowledge for acquiring, expressing, using, storing,  
retrieving knowledge, and even creating new knowledge. The major part of this level
resides within the CWM repository.

\subsubsection {XML as a Pivot Language}

The architecture we propose necessitates a universal formalism so that all its components (core, metadata, knowledge, drivers,
interface, data and knowledge interchange...) can interoperate. With its vocation for semi-structured data exchange, the eXtensible Markup Language (XML) already offers a great flexibility to represent complex data, and great possibilities for structuring, modelling, and storing them \cite{imd03}.
XML indeed allows to store together data and their description, either implicitely or through a schema definition. This type of representation is particularly useful in a data warehousing environment where such metadata are casual.
Furthermore, many XML and MOF-related facilities, such as the XML Metadata Interchange (XMI \cite{xmi05}) or the Common Warehouse Metadata Interchange (CWMI), can help in managing metadata in an XML data warehouse and specify source-type drivers, while ensuring CWM compliance.

CWM compliance is ensured by the CWM repository that is integrated into the data warehouse kernel. All $MD~\&~KB$ modules use this repository to communicate with the data warehouse. CWM, through its five metamodels (object, foundation, resource, analysis and management), provides UML components (classes, associations and packages) for modelling all the data warehouse's elements \cite{cwm01}. Table~\ref{mdkb-cwm} illustrates the correspondences between the $MD~\&~KB$ modules in our architecture and the CWM metamodels.  

\begin{table}[hbt]
	\centering
	\begin{tabular}{|l|c|c|c|c|}
		\hline
		\multirow{2}{4cm}{$MD~\&~KB$ modules} & \multicolumn{4}{|c|}{CWM metamodel} \\
		\cline{2-5}
		& Foundation & Resource & Analysis & Management \\
		\hline
		ETL / Integration & X & X & & \\
		\hline
		Administration / Monitoring & X & & & X \\
		\hline
		Analysis / Usage & X & X & X & \\
		\hline
	\end{tabular}
	\caption{$MD~\&~KB$ and CWM correspondences}
	\label{mdkb-cwm}
\end{table}

Eventually, the advances in XML warehousing \cite{P02,HBH03,RT05,adbis06bbca} render this solution plausible in the near future, especially since XML-related metadata interchange facilities integrate very well in data warehouses \cite{auth02}. Storage possibilities are also numerous, either into relational, XML-compatible DBMSs such as Oracle, SQL Server or DB2; or into XML-native DBLSs such as Lore, eXist or X-Hive. Furthermore, XML query languages such as XQuery allow the formulation of analytical queries that are intricate to express in a relational system, e.g., moving window aggregations or rollup operations on ragged hierarchies \cite{beyer05}.
Hence, our XML-based framework provides an architecture that is both extensible and ``stable", and that can be compliant with future external elements (data sources, analytical techniques and usages...). 

\section{State of the Art}
\label{sec:StateOfTheArt}

% Knowledge, Metadata, Integration -- a rediger.

Though the litterature about metadata and knowledge is abundant, the issue of integrating metadata and knowledge is scarcely addressed. In this section, we provide a quick overview of the studies that are nonetheless related to this problem. Metadata are always present in data warehouse architectures \cite{inmon02}. In our particular context, some interesting efforts aim at decentralizing the management of metadata into functional components of data warehouses \cite{mordw00,ommt05,mmde06}. They do not address the issue of domain-related knowledge, though.

Knowledge is indeed rarely exploited as such in data warehouse environments. However, issues related to knowledge management in the context of heterogeneous data warehouse environments have been addressed, by augmenting a federated warehouse with a knowledge repository \cite{kerschberg01}. Discussions about using knowledge as a basic element for managing metadata are also regularly discussed in \cite{stephens}. However, this issue is mostly addressed by the knowledge management community, which works on knowledge warehouses \cite{kw02,wecel05}, and whose focus is obviously knowledge. 

Finally, a study from the field of Geographical Information Systems (GISs, which are premium providers of complex data) is of particular interest to us. An extension of current metadata schemes has indeed been proposed to include context-based and tacit information about semantic attributes \cite{ontmd06}. These ontology-based extended metadata improve data selection and interoperability decisions. Though we are more particularly interested in explicit knowledge in our context, we can exploit this solution in our framework.

\section{Conclusion}
\label{sec:Conclusion}

In this paper, we have underlined the growing need for warehousing so-called complex data, a task that requires the management of knowledge and metadata related to these data. We enumerated the various kinds of knowledge and metadata that must be taken into account. On this basis, we proposed to integrate knowledge as metadata in the warehouse. Finally, we expanded an XML-based, CWM-compliant architecture framework for complex data warehouses we had previously proposed in the light of the new insights discussed in this paper.

One immediate perspective of our work is to validate our present proposal by experimentation, and to evaluate the impact of metadata and
knowledge integration in complex data warehouses in terms of performance. Performing performance evaluations and
comparisons, basically with and without integrating knowledge and metadata, shall show the actual relevance of our solution.

A related, important follow-up of our work is to assess the consequences of metadata and knowledge integration on traditional performance optimization
techniques such as view materialization, indexing, partitioning, query optimization, etc. These techniques will presumably need to be adapted to take into account domain-related knowledge and achieve the best performance.

Eventually, our position in this paper is to manage metadata and knowledge integration by representing knowledge
as metadata. Though we discussed arguments in favor of this particular approach in Section~\ref{sec:IntegratingKnowledgeAndMetadata}, it would be interesting to explore and assess the efficacy of the other possible solutions, namely representing metadata as knowledge or managing knowledge and metadata separately.

\bibliographystyle{lni}
\bibliography{kmeta}

\end{document}